\documentstyle[11pt,psfig,twoside]{article}

\begin{document}

\newcommand{\dd}{\,{\rm d}}
\newcommand{\ie}{{\it i.e.},\,}
\newcommand{\etal}{{\it et al.\ }}
\newcommand{\eg}{{\it e.g.},\,}
\newcommand{\cf}{{\it cf.\ }}
\newcommand{\vs}{{\it vs.\ }}
\newcommand{\zdot}{\makebox[0pt][l]{.}}
\newcommand{\up}[1]{\ifmmode^{\rm #1}\else$^{\rm #1}$\fi}
\newcommand{\dn}[1]{\ifmmode_{\rm #1}\else$_{\rm #1}$\fi}
\newcommand{\upd}{\up{d}}
\newcommand{\uph}{\up{h}}
\newcommand{\upm}{\up{m}}
\newcommand{\ups}{\up{s}}
\newcommand{\arcd}{\ifmmode^{\circ}\else$^{\circ}$\fi}
\newcommand{\arcm}{\ifmmode{'}\else$'$\fi}
\newcommand{\arcs}{\ifmmode{''}\else$''$\fi}
\newcommand{\MS}{{\rm M}\ifmmode_{\odot}\else$_{\odot}$\fi}
\newcommand{\RS}{{\rm R}\ifmmode_{\odot}\else$_{\odot}$\fi}
\newcommand{\LS}{{\rm L}\ifmmode_{\odot}\else$_{\odot}$\fi}

\newcommand{\Abstract}[2]{{\footnotesize\begin{center}ABSTRACT\end{center}
\vspace{1mm}\par#1\par
\noindent
{\bf Key words:~~}{\it #2}}}

\newcommand{\TabCap}[2]{\begin{center}\parbox[t]{#1}{\begin{center}
  \small {\spaceskip 2pt plus 1pt minus 1pt T a b l e}
  \refstepcounter{table}\thetable \\[2mm]
  \footnotesize #2 \end{center}}\end{center}}

\newcommand{\TableSep}[2]{\begin{table}[p]\vspace{#1}
\TabCap{#2}\end{table}}

\newcommand{\TableFont}{\footnotesize}
\newcommand{\TableFontIt}{\ttit}
\newcommand{\SetTableFont}[1]{\renewcommand{\TableFont}{#1}}

\newcommand{\MakeTable}[4]{\begin{table}[htb]\TabCap{#2}{#3}
  \begin{center} \TableFont \begin{tabular}{#1} #4 
  \end{tabular}\end{center}\end{table}}

\newcommand{\MakeTableSep}[4]{\begin{table}[p]\TabCap{#2}{#3}
  \begin{center} \TableFont \begin{tabular}{#1} #4 
  \end{tabular}\end{center}\end{table}}

\newenvironment{references}%
{
\footnotesize \frenchspacing
\renewcommand{\thesection}{}
\renewcommand{\in}{{\rm in }}
\renewcommand{\AA}{Astron.\ Astrophys.}
\newcommand{\AAS}{Astron.~Astrophys.~Suppl.~Ser.}
\newcommand{\ApJ}{Astrophys.\ J.}
\newcommand{\ApJS}{Astrophys.\ J.~Suppl.~Ser.}
\newcommand{\ApJL}{Astrophys.\ J.~Letters}
\newcommand{\AJ}{Astron.\ J.}
\newcommand{\IBVS}{IBVS}
\newcommand{\PASP}{P.A.S.P.}
\newcommand{\Acta}{Acta Astron.}
\newcommand{\MNRAS}{MNRAS}
\renewcommand{\and}{{\rm and }}
\section{{\rm REFERENCES}}
\sloppy \hyphenpenalty10000
\begin{list}{}{\leftmargin1cm\listparindent-1cm
\itemindent\listparindent\parsep0pt\itemsep0pt}}%
{\end{list}\vspace{2mm}}

\def\TYLDA{~}
\newlength{\DW}
\settowidth{\DW}{0}
\newcommand{\dw}{\hspace{\DW}}

\newcommand{\refitem}[5]{\item[]{#1} #2%
\def\REFARG{#3}\ifx\REFARG\TYLDA\else, {\it#3}\fi
\def\REFARG{#4}\ifx\REFARG\TYLDA\else, {\bf#4}\fi
\def\REFARG{#5}\ifx\REFARG\TYLDA\else, {#5}\fi.}

\newcommand{\Section}[1]{\section{#1}}
\newcommand{\Subsection}[1]{\subsection{#1}}
\newcommand{\Acknow}[1]{\par\vspace{5mm}{\bf Acknowledgements.} #1}
\pagestyle{myheadings}

\newfont{\bb}{timesbi at 12pt}

\begin{center}
{\Large\bf Nearby Hipparcos Eclipsing Binaries 
\vskip3pt
for Color -- Surface Brightness Calibration}  
\vskip1cm
{\bf A.~~K~r~u~s~z~e~w~s~k~i and I.~~S~e~m~e~n~i~u~k}
\vskip5mm
Warsaw University Observatory, Al.~Ujazdowskie~4,
\vskip1mm
00-478~Warszawa, Poland
\vskip1mm
e-mail: (ak,is)@astrouw.edu.pl
\end{center}

\vskip10mm

\Abstract{This paper contains the list of Hipparcos eclipsing binaries that 
fulfill the following conditions: the star is classified in the {\it Hipparcos 
Catalogue} as EA-type eclipsing binary and its parallax is either larger than 
5~mas or it is five times larger than its mean error. An eclipsing binary with 
known distance and with photometric and double-line spectroscopic orbits 
determined can be used in the process of calibrating the relation between the 
stellar surface brightness and the color that is crucial for the method of the 
distance determination by means of eclipsing binaries. This list is being 
published in order to draw attention of observers using small telescopes to 
these bright, potentially useful and in the most cases poorly observed 
objects. The advantages of the eclipsing binary method of the distance 
determination are discussed.}{binaries: eclipsing -- Stars: fundamental 
parameters -- Stars: distances} 

\vspace*{-5pt}
\Section{Introduction}
\vspace*{-3pt}
The double-line eclipsing binaries are now often considered to be one of the 
most promising distance indicators (\eg Paczy\'nski 1997). The method is 
largely geometrical with only a single relation that needs to be calibrated. 
This is the relation between the stellar surface brightness and whatever data 
that can be obtained to judge about the stellar temperature. 

The method itself is pretty much obvious once the spectroscopic and 
photometric orbits of the binary system are at hand. Yet its origins are not 
commonly known so that it happens once and again that somebody rediscovers it, 
claiming that he has found an original method. That has motivated us to write 
Section~2 where we describe how this nearly hundred years old method was 
developed. 

Section~3 is devoted to explaining why it is important to derive the 
calibration of the surface brightness -- color relation exclusively from 
observations of eclipsing binaries with known distances. 

Finally in Section~4 we present the list with a selection of nearby Hipparcos 
eclipsing binaries. 

\vspace*{-4mm}
\Section{Historical Outlook}
\vspace*{-3mm}
The photometric orbit of an eclipsing binary gives us relative radii of both 
stellar components expressed in terms of their separation and in addition an 
orbital inclination. The double-line spectroscopic orbit when combined with 
the orbital inclination that comes out from the photometric orbit results with 
masses of both components and metric value of the system dimension. 

When the distance to the binary system is known then the angular sizes of both 
components are also known and knowing the apparent magnitudes of both 
components which come out from the photometric solution makes it possible to 
calculate surface brightness for each of components. So, known parallax of an 
eclipsing binary can be used to obtain direct measurement of the surface 
brightness.  

On the other hand if one knows the value of surface brightness of an eclipsing 
binary component and if also the photometric and double-line spectroscopic 
orbits for that eclipsing binary are available then it is possible to 
calculate distance to the binary system. 

First applications of this two-way inference were made in the "distance to 
surface brightness" direction. The necessary observational data started to be 
available about hundred years ago. Vogel (1890) was first to determine radial 
velocity orbital variations for Algol and thus he obtained the first 
single-line spectroscopic orbit for any eclipsing variable. He is also 
credited with the first determination of the stellar radius, expressed in that 
case in miles. Stebbins (1910), starting observations with his newly developed 
selenium cell photometer, obtained the first accurate photometric light curve 
of Algol. Combining the photometric orbit based on his observations and the 
single-line spectroscopic orbit of Schlesinger and Curtiss (1908) and using 
average of three then existing determinations of trigonometric parallax 
(70~mas as compared with the Hipparcos value of 35~mas) he was able to 
estimate values of surface brightness for both components expressed in units 
of the solar surface brightness. With a single-line spectroscopic orbit these 
estimates depended on an assumed mass ratio of Algol components. For two 
plausible assumptions about mass ratio the resulting surface brightness 
differed by a factor of three. 

$\beta$~Aur was the first eclipsing binary with double-line spectroscopic 
orbit (Baker 1910) and with good photometric light curve (Stebbins 1911). The 
only thing that marred this otherwise excellent situation was low accuracy of 
the available data on the trigonometric parallax. Stebbins (1911) was 
analyzing this set of data under the assumption that the parallax is smaller 
than 30~mas and therefore he was able to determine only lower limits for 
surface brightness of both components. This limitation was overcome by 
Russell, Dugan and Steward (1927) who used parallax equal to 34~mas and 
obtained surface brightness of both components expressed in units of an 
equivalent effective temperature. It is worth mentioning here that the 
Hipparcos parallax for $\beta$~Aur is equal to 40~mas. 

Gaposchkin (1933) made an attempt to determine effective temperatures for 30 
eclipsing binaries with measured parallaxes even though in most cases these 
parallaxes were smaller than corresponding measurements errors. This work was 
criticized (Woolley 1934, Pilowski 1936) on obvious reason of using nonuniform 
and largely unreliable data. Kopal (1939) repeated the work of Gaposchkin 
using data on radial velocities and proper motions available for 39 systems 
and resorting to the statistical parallax method after he had divided his data 
into three groups depending on spectral types. He also used two binary systems 
with trigonometric parallaxes and two with group parallaxes. Thus before the 
year 1940 there already existed a crude independent calibration of the surface 
brightness expressed in terms of temperature as a function of spectral type, 
based exclusively on the eclipsing binaries. 

Up to now we presented the "distance to surface brightness" inference. It is 
difficult to imagine that all the involved individuals were not aware of the 
possibilities and potentials of the reverse inference "surface brightness to 
distance". In any case, we have not encountered any reference to that 
possibility prior to the papers by Gaposchkin (1938, 1940) but even in that 
case the problem of the distance determination was not stated openly. The 
luminosities of eclipsing binary components were calculated with the help of 
system dimensions and with temperatures judged from the spectral types. A 
trivial step of calculating distances by comparing luminosities with apparent 
magnitudes was not done -- as it was not done in much newer and much more 
accurate analysis by Andersen (1991). Gaposchkin stressed the fact that the 
calibration he had applied was based exclusively on the eclipsing binaries 
data. 

For nearby Galactic stars the quantities that are interesting are masses, 
sizes, luminosities and temperatures. Once we know these quantities it is not 
really relevant, if a star is 100 or 200~pc away. The situation is much 
different when we have to do with eclipsing variables in extragalactic 
nebulae. In that case it provides opportunity to determine the distance of the 
host external galaxy. Gaposchkin (1962) determined distance to an eclipsing 
variable in the M31 nebula. He did it using very crude form of the method but 
undoubtly the distance determination was the main aim of that paper and 
certainly that was the "surface brightness to distance" inference. Several 
papers with the determination of distances of eclipsing variables in M31, LMC, 
and SMC followed (Gaposchkin 1968, 1970, Dworak 1974, de Vaucouleurs 1978). 
Attention was also directed to the Galactic eclipsing binary systems. Dworak 
(1975) and Brancewicz and Dworak (1980) prepared a catalog of more than 1000 
eclipsing variables for which they made crude determination of parallaxes. 

It was a common property of both Gaposchkin and Dworak determinations of the 
distance that they did not stick to the clean case with a good photometric 
orbit and a good double-line spectroscopic orbit supplemented with information 
about temperatures of components. Eclipsing binaries offer plenty of 
opportunities for estimating the mass ratio of components in the case of 
single-line spectrum and even for estimating masses without any spectroscopic 
data. This kind of mixed accuracy data could be useful \eg for selecting 
candidates for parallax observations by Hipparcos (Dworak and Oblak 1987, 
1989) but it has not helped to the method's reputation. 

Originally it was the stellar spectral type that was used for estimating the 
temperature and consequently the surface brightness what needed also the 
knowledge of bolometric correction. Barnes and Evans (1976) found that the 
${V-R}$ color can serve as an excellent tool in that context without any need 
to know the spectral types, effective temperatures or bolometric corrections. 
All the relevant informations are compressed into so called surface brightness 
parameter $F_V$ that can be directly determined from observations. In 
particular, for stars later than the spectral type A0 the plot of surface 
brightness parameter $F_V$ \vs the ${V-R}$ color index is parallel to the 
reddening line what obviates the need for precise reddening determination. The 
Barnes--Evans finding was soon applied by Lacy (1977, 1979) to the eclipsing 
binary distance determination. 

In an early calibration Barnes, Evans and Moffett (1978) could only use three 
eclipsing binaries as calibrators, namely $\beta$~Aur, YY~Gem and CM~Dra so 
that the calibration was based mainly on stars with interferometrically 
determined angular sizes supplemented with data from lunar occultations. 
Popper (1980) has modified slightly the calibration of Barnes, Evans and 
Moffett. He allowed deviations from linearity in the relation between the 
surface brightness parameter and the color index and beside the recommended by 
Barnes and Evans ${V-R}$ index he also calibrated ${B-V}$ and Str{\"o}mgren 
${b-y}$ indices. Also separate calibrations for dwarfs and giants were given. 
Recent calibrations of the Barnes--Evans relation concerned late type stars 
(Fouqu{\'e} and Gieren 1997, Beuermann \etal 1999) or stars later than A0 
(Di~Benedetto 1998). 

The new Hipparcos data were used by Popper (1998) for comparison with his old 
(Popper 1980) calibration of the relation between the surface brightness 
parameter and the ${B-V}$ color index. He selected 14 detached eclipsing 
binaries closer than 125 pc with the mean errors of the Hipparcos parallax of 
10\% or less and with good photometric and spectroscopic data. The outcome of 
the comparison is that the majority of objects lie on or slightly above the 
calibration curve but 5 binary systems are situated clearly below it. Popper 
suggested that these 5 outliers may have depressed surface brightness due to 
spotted character of their surfaces. Ribas \etal (1998) made another selection 
of eclipsing binaries with Hipparcos parallaxes. As compared to the Popper 
selection they relaxed the distance accuracy requirement (relative errors in 
the trigonometric parallax smaller than 20\%) but sticked to the high accuracy 
of the object dimension determination. The resulting sample of 20 stars 
contains only 5 objects common with Popper sample. Ribas \etal stopped at the 
calculation of the effective temperatures for all components of these 20 
binaries and did not proceeded with collecting the color indices and 
constructing the surface brightness parameter \vs color index diagram. 
These two papers give an idea what kind of photometric and spectroscopic data 
is available right now. About ten times larger number of eclipsing variables 
have trigonometric parallaxes measured by Hipparcos with accuracy better than 
20\%, many of them discovered as eclipsing variables by Hipparcos as well, but 
majority of them lacking sufficiently good photometric and spectroscopic data.  

\vspace*{-6pt}
\Section{Motivation for Using More Eclipsing Binaries as Calibrators}
\vspace*{-3pt}
When one aims to determine accurate distances with the help of eclipsing 
binaries then the calibration of the surface brightness parameter \vs color 
index should be as good as possible and free as much as possible from any 
systematic errors. 

Angular sizes determined with the help of interferometry (Hanbury Brown \etal 
1974, Davis 1997), lunar occultations (Ridgway \etal 1980, Richichi 1977) or 
infrared flux method (Blackwell and Shallis 1977, Blackwell and Lynas-Gray 
1994) are plagued by the presence of limb darkening. They are effective sizes 
corresponding to some effective surface brightness. One can correct such 
effective sizes having some idea about the degree of limb darkening either 
from theoretical models of stellar atmospheres or from observations of limb 
darkening in eclipsing binaries. In any case the need to correct for the limb 
darkening makes the calibration less direct. When analyzing light curve of an 
eclipsing binary astronomer can determine also limb darkenings of the 
components so that the component sizes should be free from limb darkening 
uncertainty. Recent progress in the interferometric techniques opens the 
possibility to determine limb-darkened angular diameters of stars (Benson 
\etal 1997, Hummel \etal 1998, Pauls \etal 1998, Armstrong \etal 1998, Hajian 
\etal 1998) also by means of interferometry. A comparison of limb-darkening 
resulting from these two techniques can be seen as an additional cross-check 
of the calibration. 

Surface brightness dependence on gravity and metallicity is not particularly 
strong but striving for the best accuracy the corresponding corrections should 
be calibrated and applied. As these corrections are not expected to be large 
it should be enough to determine the shape of the functional dependence of 
corrections on gravity and metallicity with the help of atmospheric models but 
the zero point, or more precisely the dependence of the surface brightness 
parameter on color for solar metallicity main sequence stars, should be 
determined by comparison with the calibrating data. One of advantages of the 
eclipsing binary data is that surface gravity is also accurately known in that 
case.    

We think that the optimal case is when the eclipsing binary method of distance 
determination is calibrated exclusively with the use of eclipsing binaries 
with geometrically determined distances. This has been made feasible by 
publication of the Hipparcos trigonometric parallaxes for many nearby 
eclipsing binaries. 

Beside the use for distance determination such calibration can serve as an 
independent check for the data on fundamental stellar parameters resulting 
from other methods of stellar angular size determination including model 
calculations. 

\begin{figure}[p] 
\vspace*{2cm}
\hglue3cm\centerline{\psfig{figure=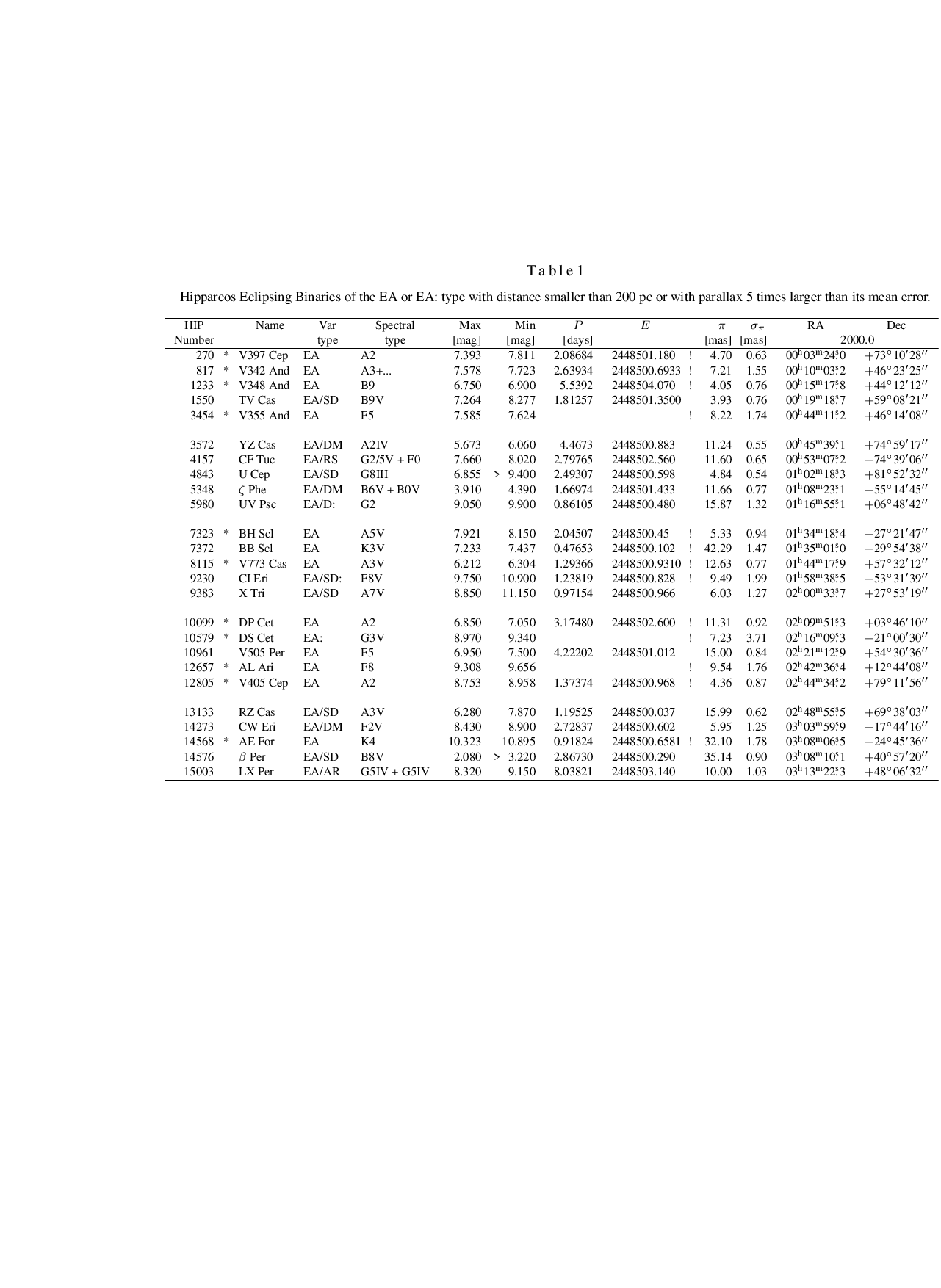,angle=90}} 
\end{figure}
\begin{figure}[p] 
\vspace*{2cm}
\hglue2cm\centerline{\psfig{figure=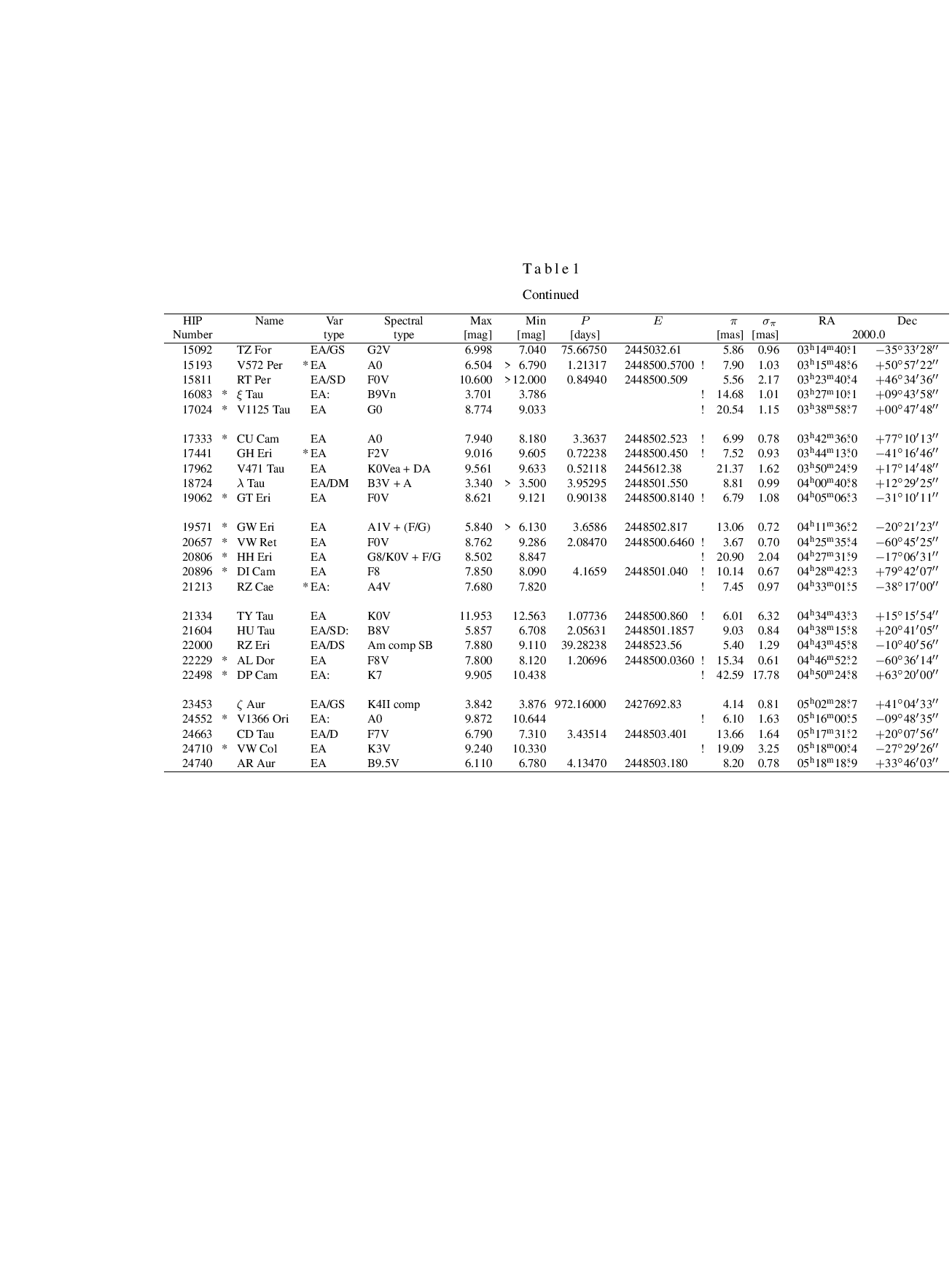,angle=90}} 
\end{figure}
\begin{figure}[p] 
\vspace*{2cm}
\hglue3cm\centerline{\psfig{figure=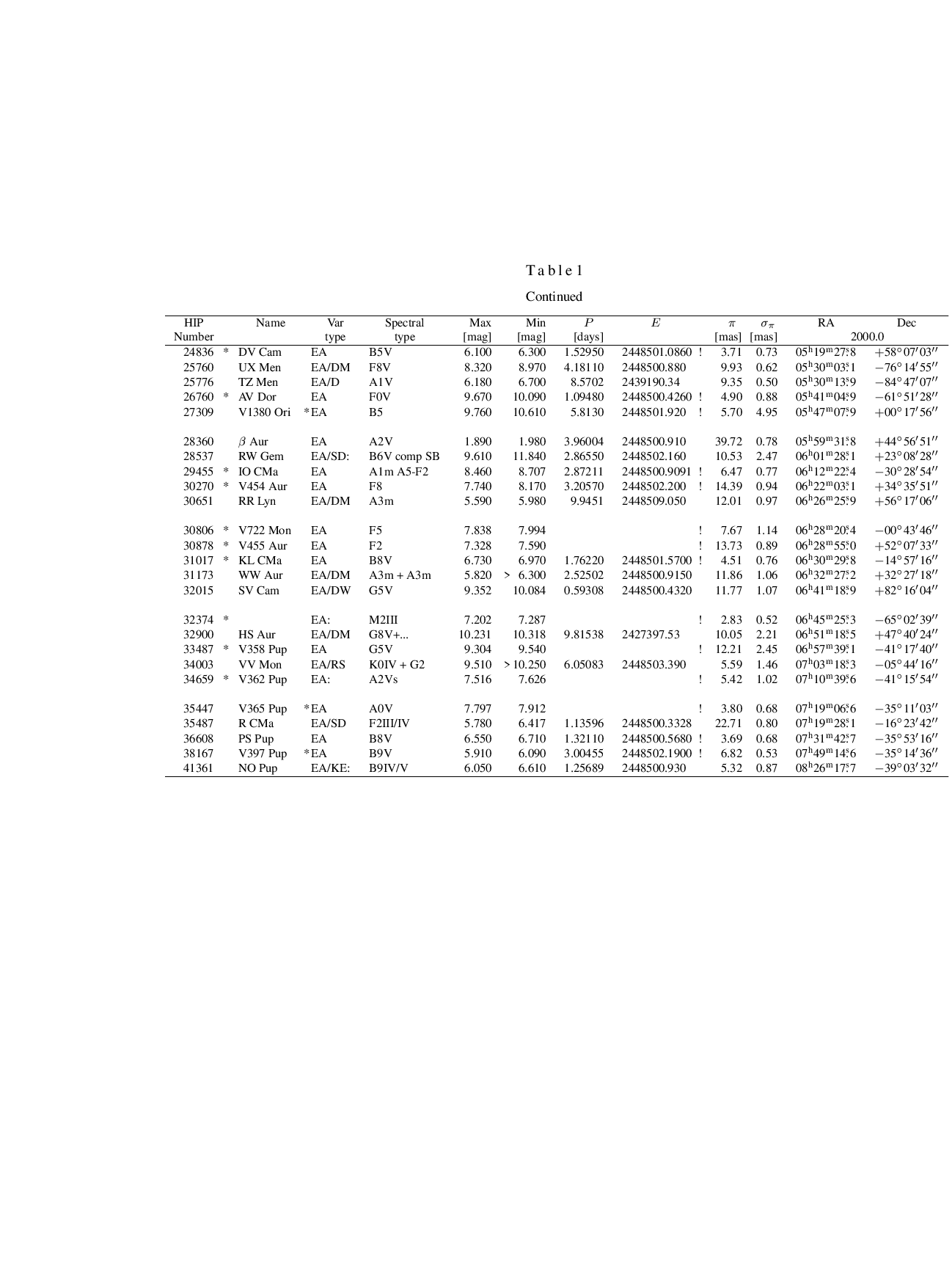,angle=90}} 
\end{figure}
\begin{figure}[p] 
\vspace*{2cm}
\hglue2cm\centerline{\psfig{figure=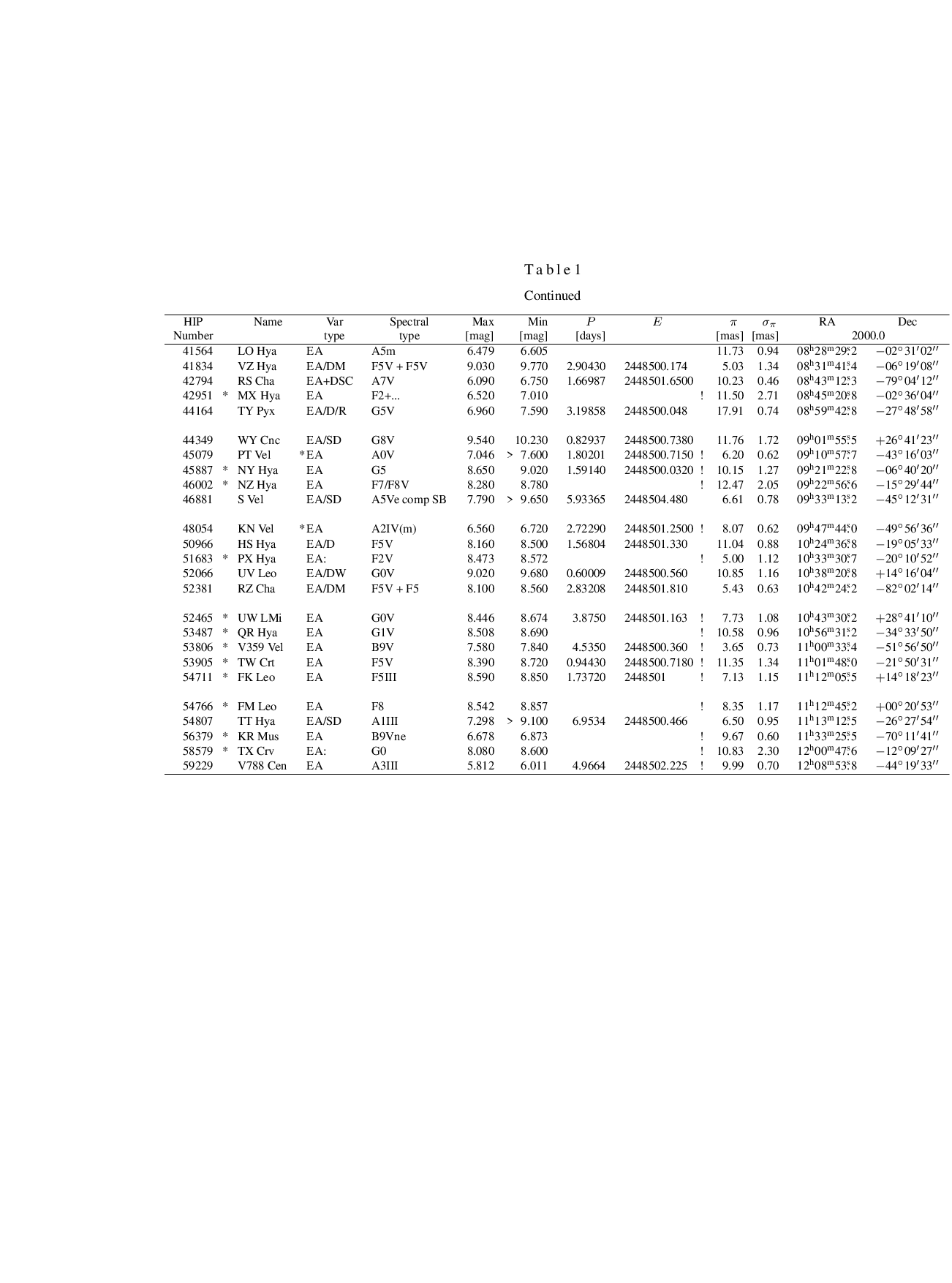,angle=90}} 
\end{figure}
\begin{figure}[p] 
\vspace*{2cm}
\hglue3cm\centerline{\psfig{figure=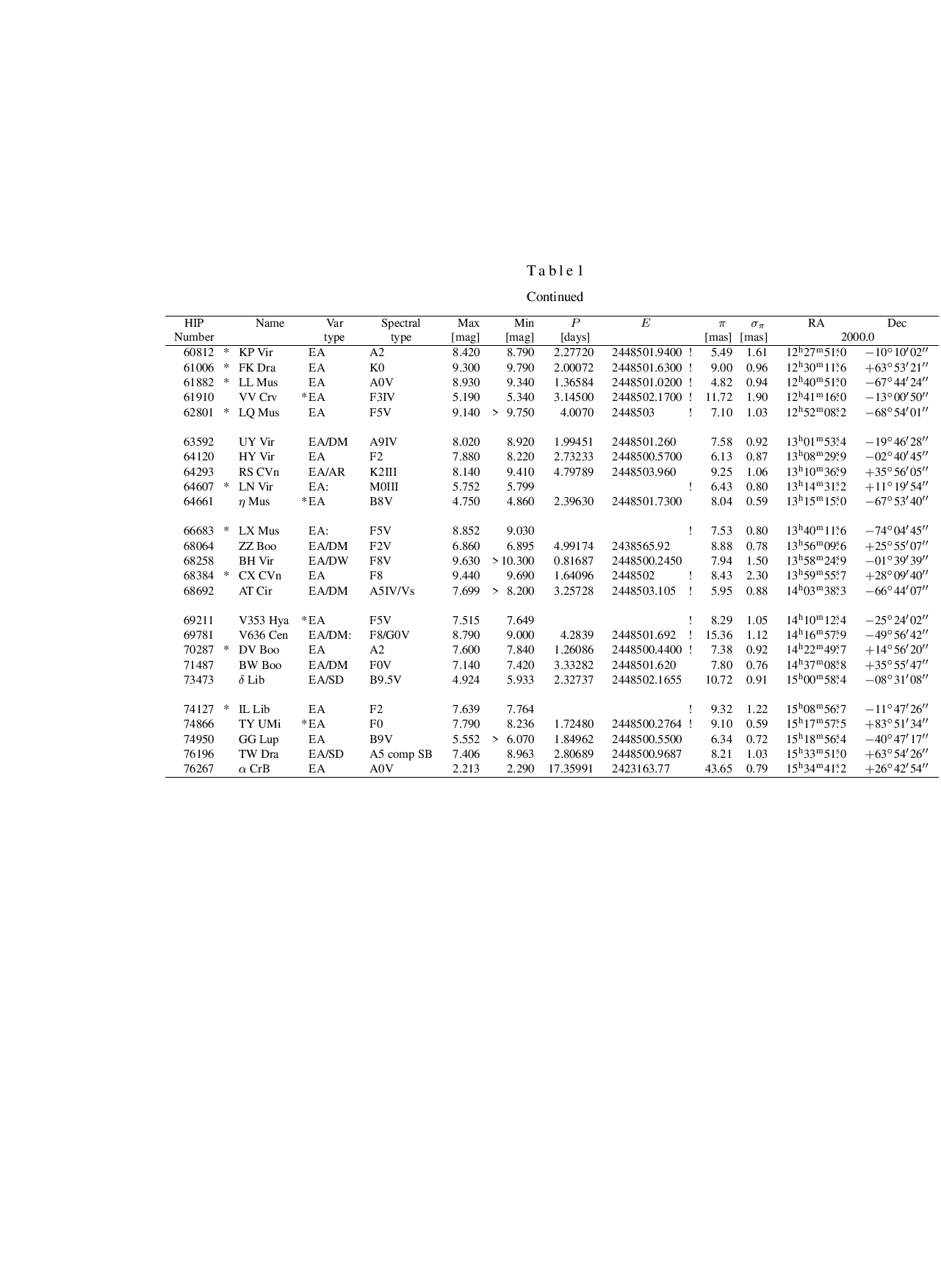,angle=90}} 
\end{figure}
\begin{figure}[p] 
\vspace*{2cm}
\hglue2cm\centerline{\psfig{figure=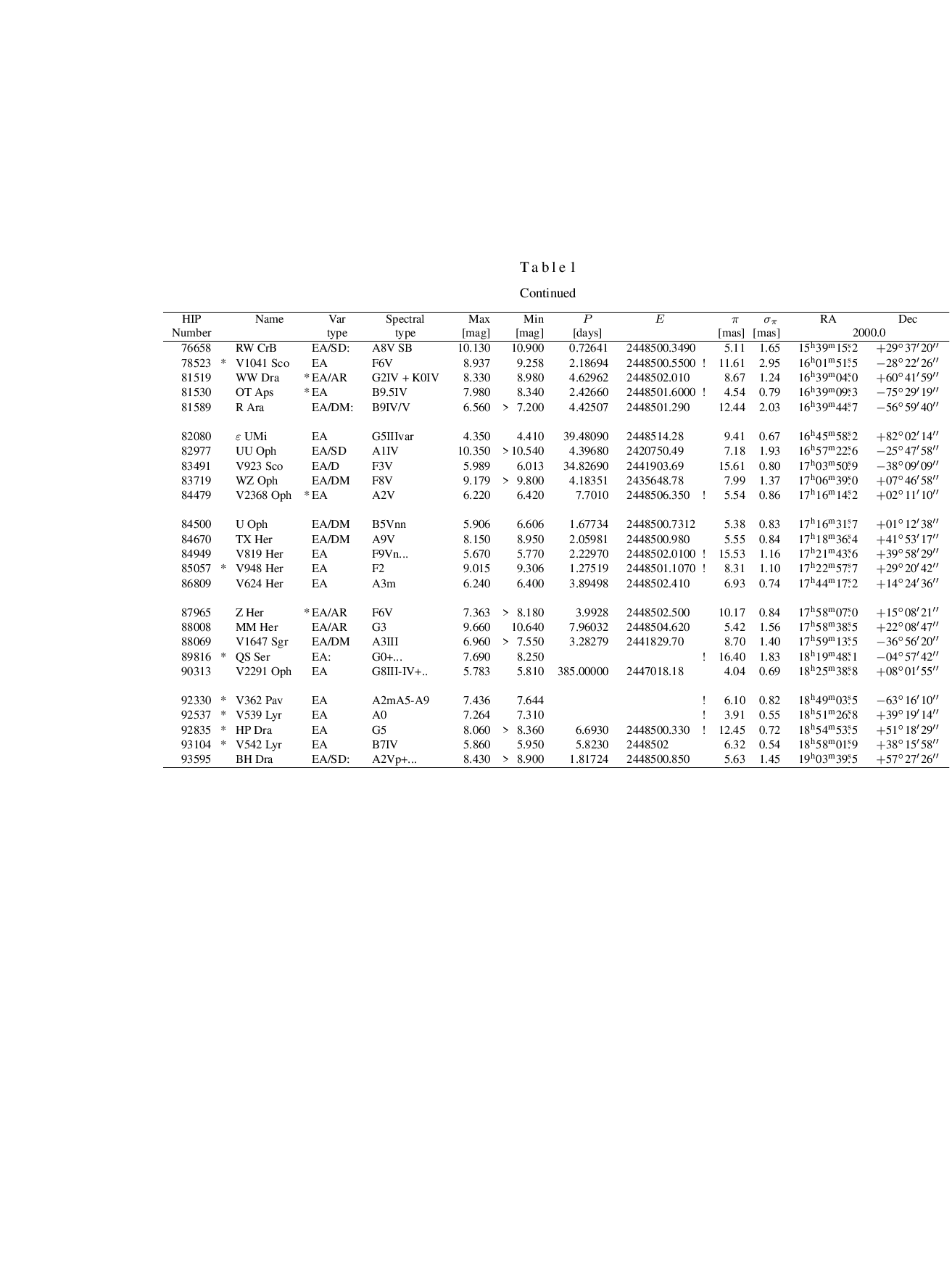,angle=90}} 
\end{figure}
\begin{figure}[p] 
\vspace*{2cm}
\hglue3cm\centerline{\psfig{figure=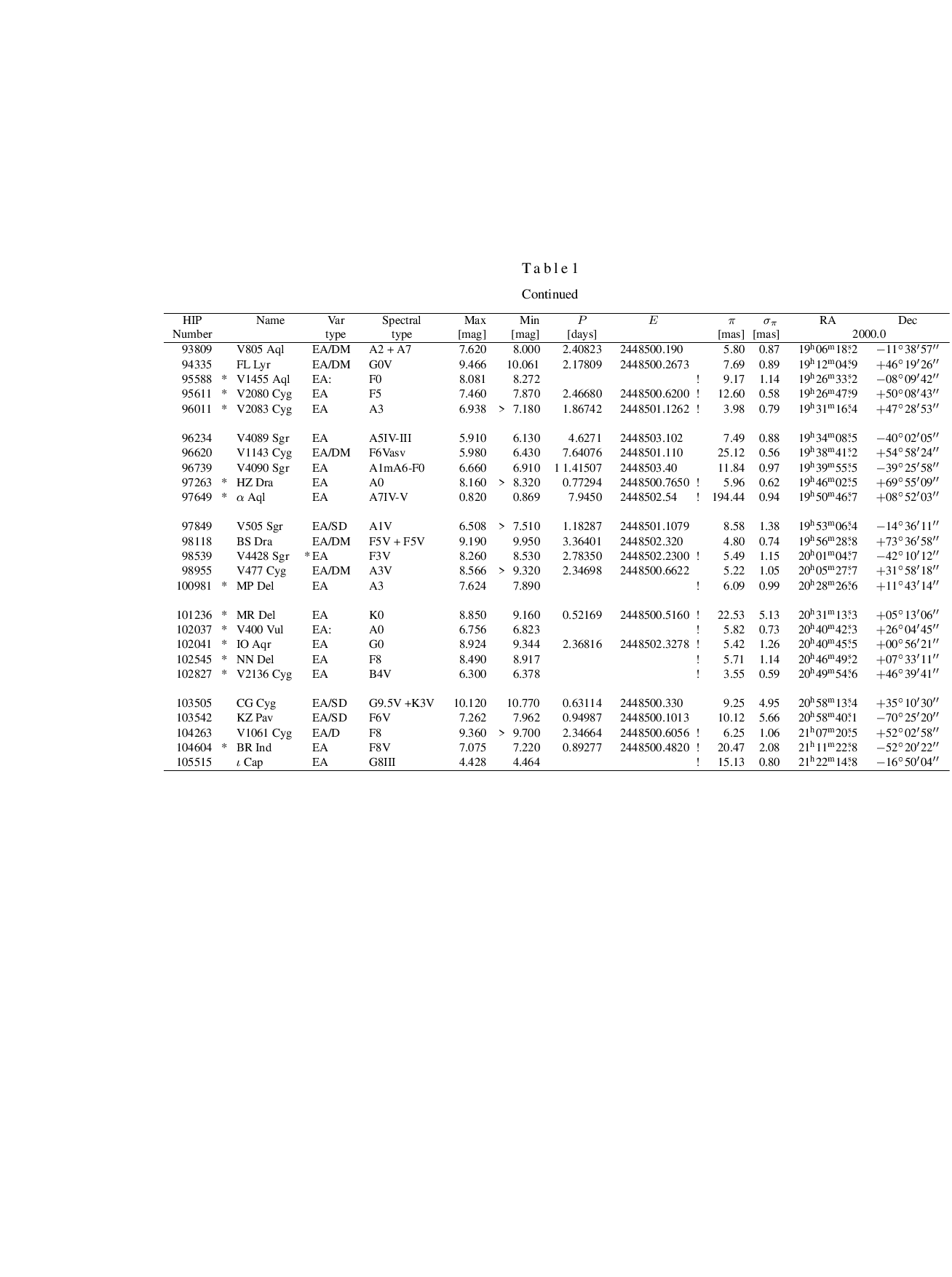,angle=90}} 
\end{figure}
\begin{figure}[p] 
\vspace*{2cm}
\hglue2cm\centerline{\psfig{figure=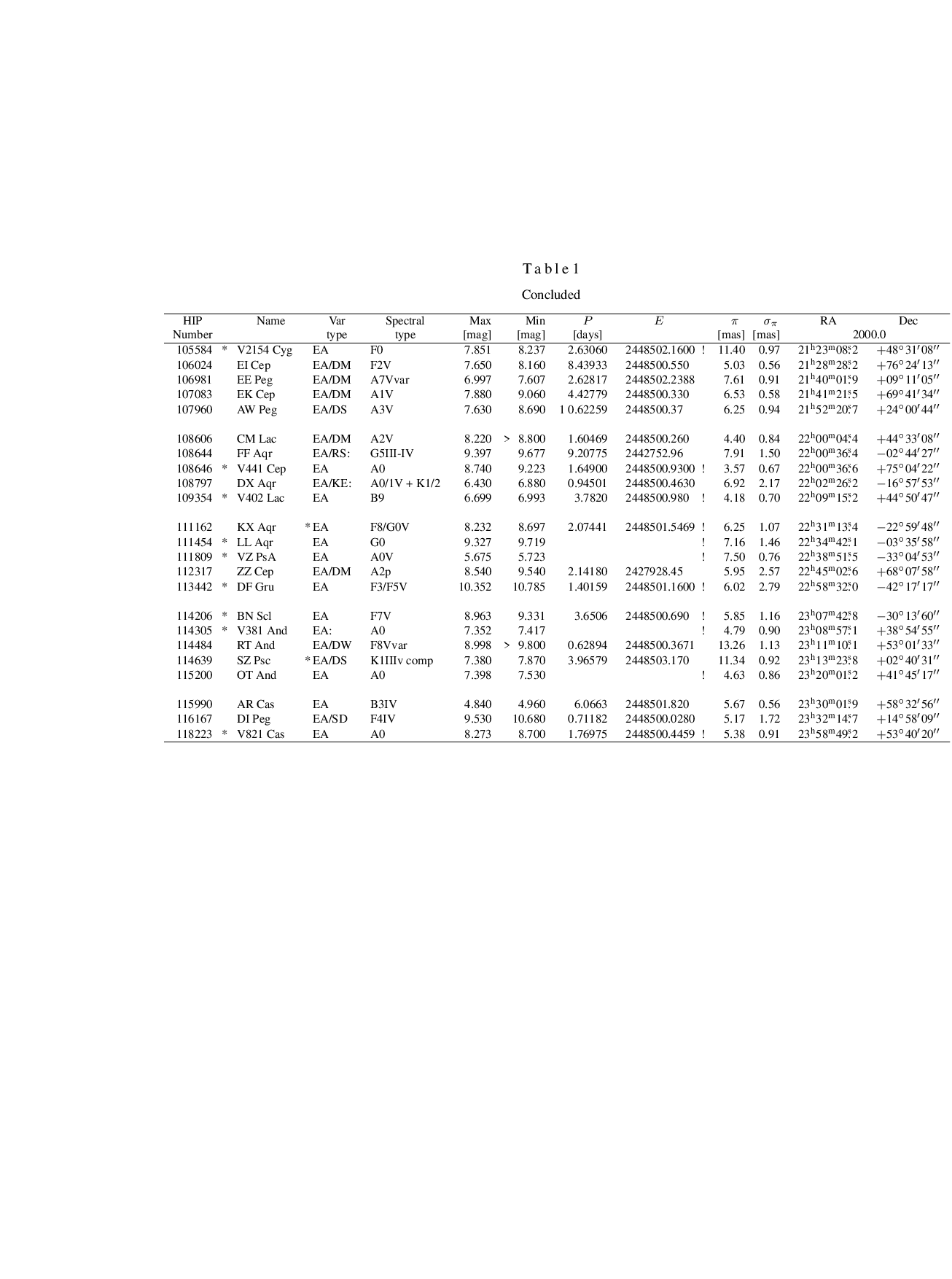,angle=90}} 
\end{figure}

In the following Section we present the list of nearby Hipparcos eclipsing 
binaries. For most of them there are only scanty observational data available. 
Some of them are not good for being reliable calibrators because of the light 
curve characteristics, RS~CVn type variability, small depths of eclipses or 
because of being semi-detached system but rejections based on such arguments 
could be done for well observed systems only. For the sake of using clearly 
defined selection criteria we have left in the Table all the objects that 
fulfill our primary criteria. 

\vspace*{-6pt}
\Section{Table Description}
\vspace*{-3pt}
Table~1 contains all Hipparcos eclipsing binaries that have their variability 
types denoted as EA or EA: and that fulfill the following distance condition: 
the binary must be either nearer than 200~pc or the standard error of its 
parallax must be five times smaller than the parallax value. In the {\it 
Hipparcos Catalogue} we have found 198 eclipsing binaries that fulfill these 
conditions. 156 of these stars are comprised in the Section "Periodic 
Variables" of the {\it Hipparcos Variability Annex} (ESA 1997, Vol.~11) and 42 
in the Section "Unsolved Variables" of this {\it Annex}. The latter Section 
contains the stars with generally unknown periods. All of them are listed in 
Table~1 in order of increasing Hipparcos numbers. The columns of Table~1 are 
generally self-explanatory. Comments must only be given to some of them. The 
asterisk between the Hipparcos number and the name of the star indicates that 
the object has been newly-classified in the {\it Hipparcos Catalogue} on the 
basis of the Hipparcos observations and the preliminary variability analysis. 
The asterisk preceding variability type in column~3 denotes that this type was 
newly classified by Hipparcos. The maximum  and minimum magnitudes in columns 
5 and 6 of the Table are taken as determined by Hipparcos. Columns 9 and 10 
give the parallax value and its standard error in milliarcseconds (mas).  

We have also selected a set of poorly observed stars that have neither 
spectroscopic nor photometric orbit solutions what has been validated by 
search in the {\sc SIMBAD} database. Such objects have been marked by 
exclamation marks between columns 8 and 9.  

\Acknow{We are greatly indebted to Professor Bohdan Paczy\'nski for continuous 
interest in this work and stimulating advices. Mr.\ Wojciech Pych is gratefully 
acknowledged for helping us at reading Hipparcos data. This work would not be 
possible without use of the {\it Hipparcos Catalogue}. Also the {\sc SIMBAD} 
database operated by the Strasbourg University was very useful. We acknowledge 
partial support from the KBN grant BST to the Warsaw University Observatory.} 

%\vbox{

%}
\end{document}